\documentclass[aps,floats,floatfix,showpacs,amssymb,prd,twocolumn,superscriptaddress,nofootinbib,nolongbibliography,10pt]{revtex4-2}

\usepackage{amssymb,amsmath,verbatim,mathtools,needspace,enumitem,etoolbox,graphicx,physics,afterpage,bigints,textcomp,gensymb,tabularx,bm,booktabs,multirow,ragged2e}

\usepackage[dvipsnames]{xcolor}
\definecolor{linkcolor}{rgb}{0.0,0.3,0.5}
\definecolor{dodgerblue}{HTML}{1E90FF}
\usepackage[unicode, colorlinks=true, linkcolor=linkcolor, citecolor=linkcolor, filecolor=linkcolor,urlcolor=linkcolor, pdfusetitle]{hyperref}
\usepackage[all]{hypcap}
\usepackage[T1]{fontenc}
\usepackage[utf8]{inputenc}
\usepackage{orcidlink}
\usepackage{tikz}

\makeatletter
\newcommand*{\balancecolsandclearpage}{\close@column@grid \cleardoublepage \twocolumngrid}
\makeatother

\usepackage{lipsum}

\newcommand{\jhu}{\affiliation{William H. Miller III Department of Physics and Astronomy,\\ Johns Hopkins University, 3400 North Charles Street, Baltimore, Maryland, 21218, USA}}
\newcommand{\GSSI}{\affiliation{Gran Sasso Science Institute (GSSI), I-67100 L’Aquila, Italy}}
\newcommand{\GranSasso}{\affiliation{INFN, Laboratori Nazionali del Gran Sasso, I-67100 Assergi, Italy}}

\begin{document}

\title{Gravitational-wave parameter estimation to the Moon and back:\texorpdfstring{\\}{}massive binaries and the case of GW231123}

\author{Francesco Iacovelli\texorpdfstring{\,}{}\orcidlink{0000-0002-4875-5862}}\email{fiacovelli@jhu.edu}
\jhu

\author{Jacopo Tissino\texorpdfstring{\,}{}\orcidlink{0000-0003-2483-6710}}
\GSSI
\GranSasso

\author{Jan Harms\texorpdfstring{\,}{}\orcidlink{0000-0002-7332-9806}}
\GSSI
\GranSasso

\author{Emanuele Berti\texorpdfstring{\,}{}\orcidlink{0000-0003-0751-5130}}%
\jhu

\pacs{}

\date{\today}

\begin{abstract}
We study the prospects of the Lunar Gravitational-Wave Antenna (LGWA), a proposed deci-Hz GW detector, to observe binary black holes (BBHs) and enable multiband science with ground-based detectors. We assess the detectability of the events observed by current instruments up to the  GWTC-4.0 data release, and of simulated populations consistent with the latest reconstruction by the LIGO--Virgo--KAGRA Collaboration. We find that LGWA alone would have been able to observe more than one third of the events detected so far, and that it could detect $\sim\!90$ events merging in the ground-based band per year out to redshifts $z\sim3-5$. Current detectors at design sensitivity and 100\% duty cycle could detect thousands of BBHs per year, with one to a few hundred multiband counterparts in LGWA. Third-generation (3G) detectors can observe most of the BBHs detected by LGWA merging in their frequency band in the simulated mass range $7\,{\rm M}_\odot\lesssim M_{\rm tot}\lesssim 600\,{\rm M}_\odot$, enabling systematic joint analyses of hundreds of events. The short time to merger from the deci-Hz band to the Hz--kHz band (typically months to a year) allows for early warning, targeted follow-up, and archival searches. Multiband observations of intermediate-mass BBHs in the deci-Hz band are particularly promising. We perform an injection study for a GW231123-like system (the most massive BBH detection to date, which accumulates $\sim\!10^5$ inspiral cycles in LGWA) and show that deci-Hz observations can measure the chirp mass even better than 3G instruments and yield good sky localization and inclination measurement, even with a single observatory. Opening the deci-Hz band would substantially improve the prospects of GW astronomy for intermediate-mass BBHs.
\end{abstract}

\maketitle

\section{Introduction}

The LIGO--Virgo--KAGRA (LVK) Collaboration~\cite{LIGOScientific:2014pky,VIRGO:2014yos,Aso:2013eba} has recently reported the observation of the most massive binary black hole (BBH) system observed to date in GWs, GW231123\_135430 (hereafter GW231123)~\cite{LIGOScientific:2025rsn}. Together with GW190521~\cite{LIGOScientific:2020iuh}, this observation confirms that intermediate-mass black hole (IMBH) binaries exist and merge in the Universe. GW231123 spent only $\sim\!5$ cycles in the LIGO detectors' frequency band, and lay in a region of parameter space particularly affected by waveform systematics, so reconstructing its properties was difficult, despite its relatively large signal-to-noise ratio (${\rm SNR}\gtrsim20$).
The total source-frame mass has been estimated to lie in the range $\sim\![190,\,265]\,{\rm M}_\odot$ with 90\% confidence. The observation of IMBH mergers is of pivotal importance for the characterization of the BBH mass spectrum. The reconstruction of features in the population, such as a putative high-mass cutoff (see e.g.~\cite{LIGOScientific:2025pvj}), does not scale linearly with the number of detectable events: massive events 
are typically more informative~\cite{Essick:2021vlx,Moore:2021xhn,Baxter:2021swn,Mancarella:2025uat}. Observing IMBH mergers could constrain current uncertainties on the pair instability supernova (PISN) mass gap (see e.g.~\cite{Farmer:2019jed,Farmer:2020xne,Woosley:2021xba}) and astrophysical models of IMBH formation~\cite{Miller:2003sc,Gair:2010dx,Veitch:2015ela,Gerosa:2017kvu,Fishbach:2017zga,2017IJMPD..2630021M,2020ARA&A..58..257G,Gerosa:2021mno,Mehta:2021fgz,Mandel:2021smh,Fragione:2022avp,Wang:2022unj,Askar:2023pmd,2025ApJ...988...15L,Kritos:2022non,Fairhurst:2023beb,Reali:2024hqf,Franciolini:2024vis,Kritos:2024upo,Kritos:2024sgd,Kritos:2025bby}.

\begin{figure*}[htb]
    \centering
    \includegraphics[width=.9\linewidth]{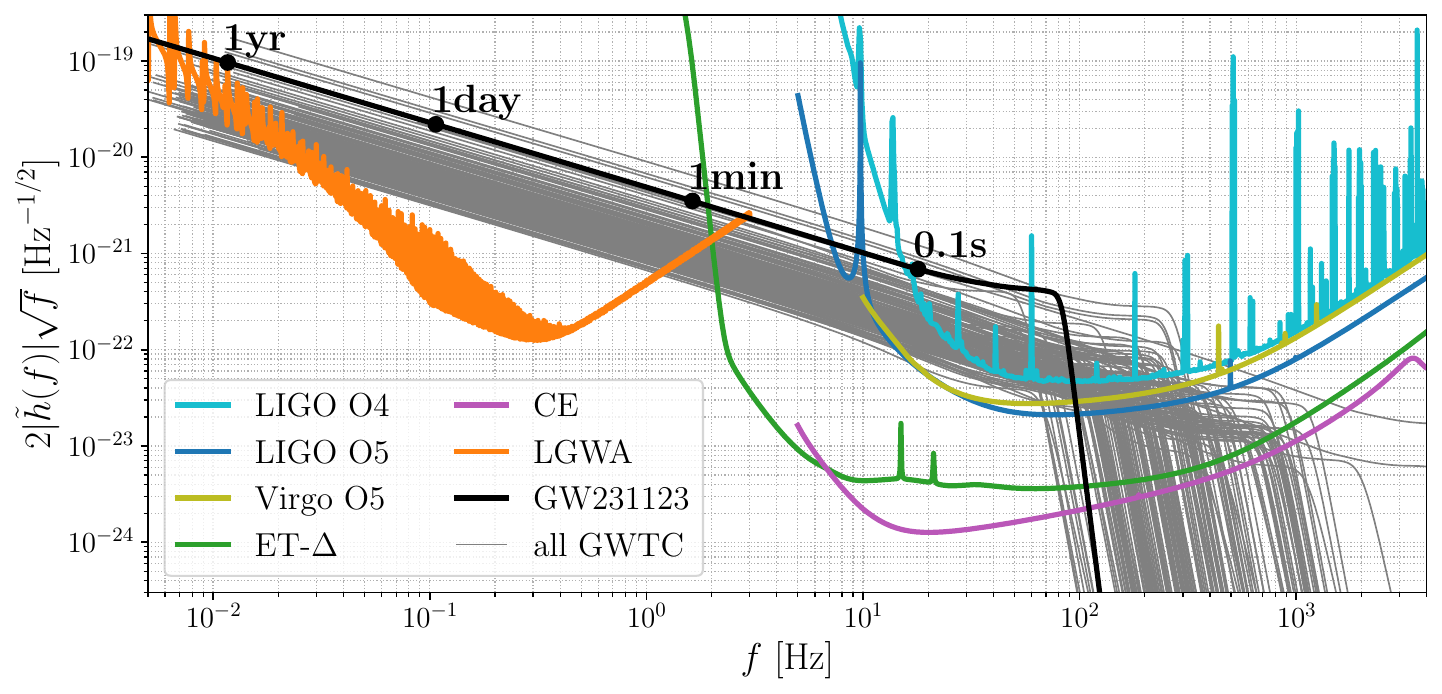}
    \caption{Sensitivity curves for the various detectors considered in this work, and representative massive BBH signals (see text).} 
    \label{fig:psds_signals}
\end{figure*}

A key ingredient to better infer the properties of these systems is the sensitivity of the detectors, particularly at low frequencies. Next-generation (3G) ground-based GW interferometers---Einstein Telescope (ET) in Europe~\cite{Punturo:2010zz,Hild:2010id,Maggiore:2019uih,Branchesi:2023mws,ET:2025xjr} and Cosmic Explorer (CE) in the US~\cite{Reitze:2019iox,Evans:2021gyd,Evans:2023euw}---will bring drastic improvements in sensitivity (see Fig.~\ref{fig:psds_signals}). For ET, with projected sensitivity down to $\sim\!2\,{\rm Hz}$, a GW231123-like signal would be observable for $\gtrsim 30$\,s as opposed to the $\sim\!0.1\,{\rm s}$ in the LIGO band, resulting in $\gtrsim 100$ observable inspiral cycles. Detectors in the deci-Hz band would be even better suited to the observation of massive BBH systems.
Proposed experiments covering this band, that would bridge the gap between ground-based detectors and LISA~\cite{LISA:2017pwj,LISA:2024hlh}, include space-borne detectors such as DECIGO~\cite{2006CQGra..23S.125K,Kawamura:2020pcg}, atom interferometers in space~\cite{2016PhRvA..94c3632H,Graham:2016plp}, lunar interferometric detectors such as the Laser Interferometer Lunar Antenna (LILA)~\cite{Jani:2020gnz,Jani:2025uaz}, and 
experiments aiming at using the Moon as a planetary-scale GW antenna monitored by seismic sensors such as the Lunar Gravitational-Wave Antenna (LGWA)~\cite{LGWA:2020mma,Ajith:2024mie}, taking advantage of the low seismic noise of the lunar environment inferred from the Apollo data~\cite{1970GeCAS...1.2309L}. Massive BBH observations are one of the main scientific objectives of deci-Hz detectors~\cite{Sedda:2019uro,Ajith:2024mie}: for example, a GW231123-like system starting at $0.1$\,Hz would be observable for 28 hours or ${\cal O}(10^4)$ inspiral cycles before merger, allowing for exquisite parameter estimation (PE).

Motivated by these observations, in this paper we explore the prospects of observing massive BBHs with (\emph{i}) current second-generation (2G) detectors, (\emph{ii}) future ground-based 3G detectors, and (\emph{iii}) the LGWA~\cite{LGWA:2020mma,Ajith:2024mie}. In particular, we perform Bayesian PE for a GW231123-like system to quantify how well the LGWA could measure its properties, and how it would complement present and future ground-based observatories.

\section{Analysis setup}

\subsection{Detectors and sensitivity}\label{subsec:detectors}

In our analysis we will include the following ground-based detector networks: 

\noindent
{\bf LIGO O4:} The Hanford and Livingston LIGO interferometers at the time of the detection of GW231123~\cite{LIGOScientific:2025snk}. In Fig.~\ref{fig:psds_signals}, for reference, we report the sensitivity of the Livingston detector.

\noindent 
{\bf LIGO-Virgo O5:} A network comprising the two LIGO interferometers and Virgo at the projected sensitivities for the fifth observing run (O5)~\cite{KAGRA:2013rdx}.

\noindent 
{\bf ET-$\Delta$:} A single triangular ET detector located in Sardinia with 10\,km arms, consisting of three nested detectors with $60^\circ$ opening angle; each of them features two instruments, one optimized for low-frequency sensitivity, and one for high-frequency sensitivity. We adopt the sensitivity curve used in Ref.~\cite{Branchesi:2023mws}.

\noindent 
{\bf ET-$\Delta$ + CE:} A network comprising ET in the triangular configuration as well as a 40\,km arm CE detector located in Idaho, US~\cite{Borhanian:2020ypi}, with the same sensitivity curve as in Refs.~\cite{Srivastava:2022slt,CEsensitivity}.

For the position and orientation of the detectors, we refer the reader to Ref.~\cite{Iacovelli:2022mbg}. Here we do not consider the ET geometry consisting of two separated L-shaped detectors in Europe~\cite{Branchesi:2023mws}, but we expect that, for the case of the GW231123-like injection studied in Sec.~\ref{subsec:pe_231123}, it would yield PE capabilities comparable to the triangular configuration for the binary's intrinsic parameters~\cite{Branchesi:2023mws,ET:2025xjr}, and we have verified through a \texttt{BAYESTAR}~\cite{Singer:2015ema} injection that the sky location posterior would show a multimodal structure, albeit with less separated modes compared to the triangular geometry. In such geometry consisting of 2 instruments, this can be traced to (\emph{i}) the used separation of $\sim\!1.1\times10^3\,{\rm km}$ between the two L-shaped ETs~\cite{Branchesi:2023mws}, and (\emph{ii}) the low merger frequency of the event and the short signal duration (see Refs.~\cite{Fairhurst:2009tc,Fairhurst:2010is,KAGRA:2013rdx} and Refs.~\cite{Santoliquido:2025aiq,Iacovelli:2026ixl} for a discussion in the context of current and future instruments, respectively).

\noindent
{\bf LGWA:} The LGWA design consists of an array of four seismometers, as described in Refs.~\cite{LGWA:2020mma,Ajith:2024mie}. The relative distance between the seismometers can be considered negligible in the present context, but the real configuration will achieve active noise cancellation by having three outer seismometer forming a triangle and a fourth seismometer in the center.
We employ the more optimistic (``silicon'', Si) model of the detector noise curve (publicly available as part of the \texttt{GWFish} package~\cite{Dupletsa:2022scg}), with low readout noise and high quality factor. The assumed mission duration is 10\,yr, and the detector is placed at a latitude $5^\circ$ above the lunar south pole, with alignment defined with respect to the Moon principal axis system, as described in Ref.~\cite{Park_2021_ephem}. 

\noindent We assume a 100\% duty cycle for ground-based detectors and for the LGWA. 

To highlight the complementarity of LGWA with ground-based interferometers, in Fig.~\ref{fig:psds_signals} we plot the sensitivity curves as well as (in gray) the waveform amplitudes $|\tilde{h}(f;\bm{\theta})| = |\tilde{h}_+(f;\bm{\theta}) + i\tilde{h}_\times(f;\bm{\theta})|$ 
(not projected onto any detector) of all the signals observed to date by LVK. These were computed using the \textsc{IMRPhenomXAS} waveform model for the dominant emission mode with aligned spins~\cite{Pratten:2020fqn} evaluated at parameters corresponding to the maximum likelihood of the public PE samples~\cite{LIGOScientific:2018mvr,LIGOScientific:2020ibl,KAGRA:2021vkt,LIGOScientific:2025slb}, and truncating each waveform 10~years before merger. The signal from GW231123 is highlighted in black, with black circles marking the frequencies that correspond to selected times to merger.

\subsection{Modeling of the LGWA response}

A key ingredient of our analysis is the modeling of the LGWA response function. Following Ref.~\cite{Dupletsa:2022scg}, the response tensor $\cal A$ of a single LGWA seismometer is given by the tensor product of the unit vector along the measurement direction tangential to the Moon surface, $\hat{\bm{e}}_x(t)$, with the unit vector normal to the surface itself, $\hat{\bm{e}}_n(t)$:
\begin{equation}
    {\cal A}(t) = \hat{\bm{e}}_x(t) \otimes \hat{\bm{e}}_n(t)\,.
\end{equation}
Moreover, each seismometer will measure displacements in two orthogonal directions, which we consider independent. Thus, a single seismometer is effectively treated as two detectors oriented at $90^\circ$ with respect to each other. 

Given the length of the signals from typical LIGO-Virgo binaries in the LGWA band, the detector response must take into account the amplitude and phase modulations as well as the Doppler term due to the Moon's motion during the observation time~\cite{Cutler:1997ta,Cornish:2003vj}. The modulations include the motion of the Earth-Moon system around the Sun (with a period of 1\,yr) as well as the motion of the Moon around the Earth and the rotation of the Moon on its axis (both with a period of $\sim\!27.3$\,days). We work in the frequency domain and implement these modulations within the stationary phase approximation (SPA), that is valid if the change in the amplitude during a cycle is much slower than the corresponding change in the phase (see e.g.~\cite{Iacovelli:2022bbs}). 
We consider only the leading-order contribution of the modulations to the SPA signal, i.e., we compute the stationary point $t^*$ as a function of frequency. Higher-order corrections, expected to be subdominant given the frequency of the modulations and the detector sensitivity band, could be computed as done for the LISA case in Ref.~\cite{Marsat:2018oam}; while this may be important when dealing with real data in the future, we do not expect it to significantly affect our results.
Then the amplitude modulation implies that ${\cal A}(t)$, and thus the detector ``pattern functions,'' become a function of time given by
\begin{equation}\label{eq:t_of_f}
    \begin{aligned}
        t^*(f) \simeq t_c - \dfrac{5}{256} &\left(\dfrac{G{\cal M}_c}{c^3}\right)^{-5/3} (\pi f)^{-8/3} \\
        &\times \left[1 + {\cal O}(\pi M_{\rm tot} f G /c^3)^{2/3}\right]\,,
    \end{aligned}
\end{equation}
where $t_c$ is the time of coalescence, $M_{\rm tot}=m_1+m_2$ is the detector-frame total mass of the binary, ${\cal M}_c = (m_1 m_2)^{3/5}/(m_1+m_2)^{1/5}$ is the detector-frame chirp mass, $m_1$ and $m_2$ are the detector-frame component masses with $m_1\geq m_2$, and we include terms up to 3.5PN order, as in Ref.~\cite{Buonanno:2009zt}. To convert the sky coordinates (right ascension $\alpha$ and declination $\delta$) in the geocentric International Celestial Reference System (ICRS) frame to the time-dependent Moon frame of the detector we use the libration angles obtained from the ephemeris as implemented in the \texttt{PyEphem} package~\cite{2011ascl.soft12014R}, using data from release DE440~\cite{Park_2021_ephem} of the NASA JPL SPICE system~\cite{ACTON199665,2018P&SS..150....9A}. 
The phase modulation arising from the time-dependent pattern functions is implicit if the signal is separated in its two polarizations: 
\begin{equation}\label{eq:signal_pol}
    \tilde{s}(f;\bm{\theta}) = \tilde{h}_+(f;\bm{\theta}) F_+ (t^*(f);\bm{\theta}) + i\tilde{h}_\times(f;\bm{\theta}) F_\times (t^*(f);\bm{\theta})\,,
\end{equation}
where $\bm{\theta}$ denotes the parameters of the event, and we explicitly write the dependence of the pattern functions $F_{+},\,F_{\times}$ on the stationary point.

The contribution of the Doppler effect to $t^*(f)$ and to the amplitude of the Fourier-transformed signal is completely negligible, because the Doppler modulation frequencies are much smaller than the frequencies relevant for LGWA. However, the Doppler phase contribution must be taken into account, because it gives rise to a time (and frequency) dependence of the time delay corresponding to
the travel time of the signal from the origin of the reference frame to the detector:
\begin{equation}
    \begin{aligned}
        \phi_L (t^*(f);\bm{\theta}) &= 2\pi f \Delta t_L (t^*(f);\bm{\theta}) \\
        &= -2\pi f \hat{\bm{s}}(\alpha,\delta) \cdot \bm{d}(t^*(f)) / c\,,
    \end{aligned}
\end{equation}
where $\hat{\bm{s}}$ denotes the unit vector pointing towards the source, and $\bm{d}$ is the vector connecting the origin of the reference frame to the detector, 
computed once again using the position ephemeris from \texttt{PyEphem}. 

As our analyses with 3G detectors use a geocentric frame, we do the same for LGWA so that the results are directly comparable. However, this means that we implicitly assume the geocenter to be stationary (or moving with constant velocity) with respect to the source during the observation period.
The validity of this approximation depends on the length of the signal in band, i.e., on the significance of the Doppler shift due to the accelerated motion of the Earth-Moon system around the Sun.

As Doppler shifts only impact phase, this choice has no impact on our population studies.
For our analysis of GW231123, most of the SNR is accumulated in the last days of observation, so we expect this choice to have a small impact. This can induce a potential bias and lead to slightly underestimate the localization capability of the LGWA. 

We implement these projections as part of the public package \texttt{gwfast}~\cite{Iacovelli:2022mbg}. 
We cross-checked our likelihood against an independent implementation that uses a reference frame comoving with the Solar System Barycenter, discussed in Ref.~\cite{Tissino_inprep}. 

\subsection{Data, approximants and sampling}\label{subsec:data}

To forecast the capability of LGWA of observing the signals from the first four Gravitational-Wave Transient Catalogs (GWTCs), we download public posterior samples from the respective \textsc{Zenodo} entries~\cite{GWTC2PE,GWTC3PE,GWTC4PE}. The latest catalog has a total of 176 events, with the maximum-likelihood waveforms (computed up to 10\,yr before merger) shown in Fig.~\ref{fig:psds_signals}. For each event, we use the \textsc{Mixed} posterior samples. For the analysis of GW231123 we consider the samples obtained with the \textsc{IMRPhenomXPHM} approximant~\cite{Pratten:2020ceb}---the same used to analyze this specific event as observed by  different detectors. This approximant models the emission from the dominant $(2,2)$ mode of the signal along with the $(2,1),\,(3,3),\,(3,2),\,(4,4)$ subdominant harmonics, and models also spin precession. 

Notice that we use the time-frequency correspondence of Eq.~\eqref{eq:t_of_f} also for higher-order harmonics of the signal. It would be more accurate to take into account that each harmonic mode will have a different relation, with the time spent in band for the generic $(\ell,\,m)$ mode being $\sim(m/2)^{8/3}$ times longer than the dominant $(2,2)$ mode: see e.g.~\cite{Kapadia:2020kss}. However, the ratio of the $(2,2)$ mode SNR to the total SNR in LGWA is ${\rm SNR}_{22}^2/{\rm SNR}^2 \simeq 99.8\%$ for GW231123. We also find ${\rm SNR}_{22}^2/{\rm SNR}^2 \geq 94.9\%$ for the full set of simulated binaries, so the impact of this approximation is negligible.

Each event is described by 15 parameters
\begin{equation}
    \begin{aligned}
        \bm{\theta} = \{&{\cal M}_c,\eta,d_L,\alpha,\delta,\theta_{\rm JN}, \psi,t_{c},\Phi_c,\chi_1,\chi_2,\theta_1, \theta_2,\phi_{12},\phi_{\rm JL}\}\,,
    \end{aligned}
\end{equation}
where ${\cal M}_c$ is again the (detector-frame) 
chirp mass, $\eta=(m_1 m_2)/(m_1+m_2)^{2}$ is the symmetric mass ratio, $d_L$ is the luminosity distance, $\theta_{\rm JN}$ is the angle between the binary's total angular momentum and the line of sight, $\psi$ is the polarization angle, $\Phi_c$ is the coalescence phase, $\chi_1$ and $\chi_2$ are the spin magnitudes of the two objects, $\theta_1$ and $\theta_2$ are the corresponding spin tilt angles, $\phi_{12}$ is the difference in azimuthal angles between the two spins, and $\phi_{\rm JL}$ is the azimuthal angle between the total and orbital angular momentum of the binary. 

{%
\setlength{\tabcolsep}{11.5pt}
\begin{table}[tb]
    \centering
    \begin{tabular}{ccc}
    \toprule\midrule
    Parameter & Value & Prior range \\
    \midrule\midrule
        ${\cal M}_c$ & $119.9\,{\rm M}_\odot$ & $[70,\,160]\,{\rm M}_\odot$\\
        $\eta$ & 0.236 & $[0.16,\, 0.25]$\\
        $d_L$ & 0.850\,Gpc & $[0.1, 2.6]\,{\rm Gpc}$\\
        $\alpha$ & 3.33 & $[0,\, 2\pi]$ \\
        $\delta$ & 0.36 & $[0,\,\pi]$ \\
        $\theta_{\rm JN}$ & 1.57 & $[0,\,\pi]$\\
        $\psi$ & 2.64 & $[0,\,\pi]$ \\
        $t_c$ & 1384722882.6\,s & $[t_c - 0.2,\, t_c + 0.2]\,{\rm s}$\\
        $\Phi_c$ & 5.03 & $[0,\,2\pi]$\\
        $\chi_1$ & 0.79 & [0,\,1]\\
        $\chi_2$ & 0.85 & [0,\,1]\\
        $\theta_1$ & 1.81 & $[0,\,\pi]$\\
        $\theta_2$ & 0.85 & $[0,\,\pi]$\\
        $\phi_{12}$ & 4.39 & $[0,\,2\pi]$\\
        $\phi_{\rm JL}$ & 1.70 & $[0,\,2\pi]$\\
        \midrule\bottomrule
    \end{tabular}
    \caption{Injection parameters and prior ranges used for the analysis of GW231123. All the angles are given in radians.}
    \label{tab:231123_params_and_priors}
\end{table}
}%

For the GW231123-like injection, we use the parameters and prior ranges listed in Table~\ref{tab:231123_params_and_priors}. For consistency with the LVK analysis, the chosen mass prior is uniform in the component masses, the prior on $d_L$ is uniform in comoving volume and source-frame time, the prior on $\delta$ is uniform in sine, the prior on $\theta_{\rm JN}$ and $\theta_{1,2}$ is uniform in cosine, and the priors on $\alpha$, $\Phi_c$, $\psi$, $\phi_{12}$ and $\phi_{\rm JL}$ are periodic. The reference frequency for the analysis is set to 10\,Hz. The time used for the analysis is a few hours off from the actual detection time in LIGO, as this yields a slightly higher SNR for the event in LGWA.
We analyze this GW231123-like event with \texttt{parallel bilby}~\cite{Smith:2019ucc} for ground based detectors, and with \texttt{bilby}~\cite{Ashton:2018jfp} (sampling a likelihood obtained through \texttt{gwfast}) for LGWA. In each case we use the \texttt{dynesty}~\cite{Speagle:2019ivv} nested sampler with 2048 live points and the \textsc{acceptance-walk} scheme, and the LIGO Algorithm Library (\texttt{LAL})~\cite{2020ascl.soft12021L}  implementation of \textsc{IMRPhenomXPHM}.

For our population analysis we consider the latest LVK population results~\cite{LIGOScientific:2025pvj}, adopting the \textsc{Broken Power Law + 2 Peaks} model for the source-frame individual masses, \textsc{Gaussian Component Spins} for the spin magnitudes and tilts, and \textsc{Power Law Redshift} for the redshift. We use the public posterior samples for the hyperparameters describing those distributions available on \textsc{Zenodo}~\cite{GWTC4Pop}. We extend the redshift distribution to high redshift by using a Madau-Dickinson profile~\cite{Madau:2014bja} in which the low-$z$ slope is fixed to the value estimated by the LVK Collaboration, and the other parameters assume typical values (i.e., a peak redshift of $z_p=2$ and a high-$z$ slope $\beta_z=3$~\cite{Madau:2016jbv}).

\section{Results and discussion}

\subsection{Prospects of observing BBHs with LGWA}

We start by assessing the capability of LGWA of observing BBHs that merge in the ground-based band, extending the analysis of Ref.~\cite{Ajith:2024mie} (see also Ref.~\cite{Song:2025lpa,Dong:2025ikq,Singh:2025cky} for joint intermediate-mass black hole (IMBH) detection prospects,  Ref.~\cite{Yelikar:2025jwh} for joint observations of a GW170817-like signal, and Ref.~\cite{Benetti:2025rxe} for a discussion of double white dwarfs).

We first consider the public posterior samples of all the events detected to date by the LVK Collaboration, as  discussed in Sec.~\ref{subsec:data}. For each event, we extract 1000 samples from its posterior, and compute the LGWA SNR of those synthetic events. We find that, among the 176 events in the catalog, LGWA would have been able to observe 
\begin{equation*}
    N_{\rm GWTC} = 56\substack{+9\\-5} \ \left(100\substack{+10\\-8}\right)
\end{equation*}
\begin{figure}[htb]
    \centering
    \includegraphics[width=\linewidth]{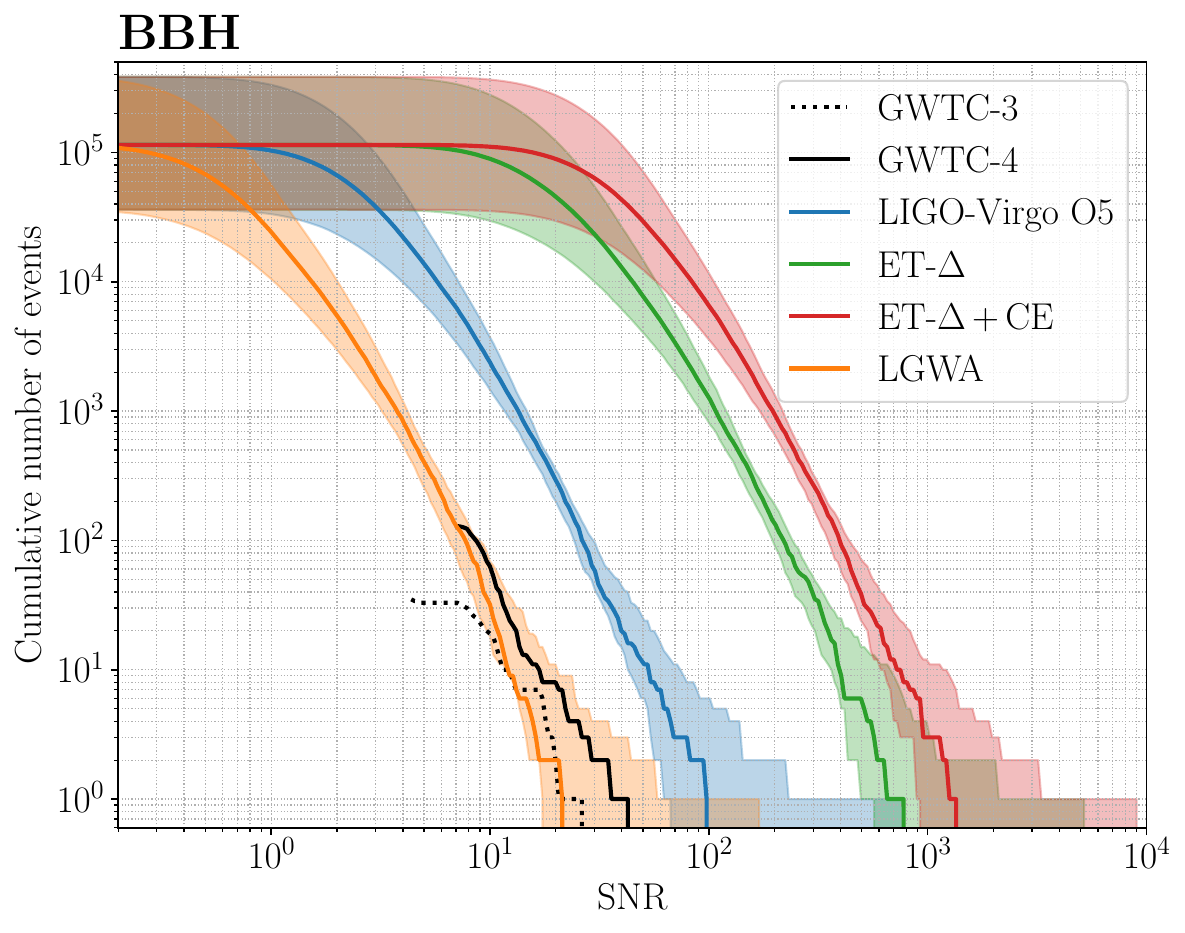}
    \caption{Inverse cumulative distribution of the number of sources as a function of the SNR for the detectors (and detector networks) considered in this work. The solid lines are the cumulative distributions corresponding to the maximum likelihood estimated for the reference population model used in Ref.~\cite{LIGOScientific:2025pvj}, while the shaded bands enclose the maximum and minimum value across 1000 simulated catalogs with varying hyperparameters. For reference, we further report the cumulative distribution of the SNRs for the sources in GWTC-3~\cite{KAGRA:2021vkt} and GWTC-4~\cite{LIGOScientific:2025slb}.}
    \label{fig:cumulative_SNR_gwtcpop}
\end{figure}

\noindent events with SNR larger than 8 (5). An SNR of 8 represents a suitable detection threshold, while an SNR of 5 has been identified as a viable threshold for archival searches of events detected with ground-based observatories in the case of LISA~\cite{Wong:2018uwb,Toubiana:2022vpp}. The uncertainty in the number of detections is a consequence of the uncertainty in the events' parameters reconstruction.  

Clearly this is not the total number of signals observable with LGWA in the mass range accessible to ground-based detectors, but only the subset of events that would have been detected in conjunction with LIGO-Virgo until the end of O4a. However, this estimate is already an indication of the high complementarity and of the promising multiband capabilities enabled by combining deci-Hz experiments with ground-based observations. By contrast, joint multiband observations between ground-based detectors and LISA are less encouraging~\cite{Gerosa:2019dbe,Moore:2019pke,Klein:2022rbf,Toubiana:2022vpp,Buscicchio:2024asl}. As suggested in Ref.~\cite{Ajith:2024mie} (and better quantified below), multiband detections lead to improved PE by combining the long signal duration in the deci-Hz band with the complementary information enabled by merger observations in the Hz-kHz band. 

To characterize more broadly the number of detectable events we adopt a different approach. Assuming the reference model used in the latest LVK population analysis~\cite{LIGOScientific:2025pvj} (see Sec.~\ref{subsec:data}), we extract $10^3$ samples from the population (hyper-)posterior, and use each sample to simulate a corresponding individual realization of a BBH catalog for 1\,yr of observations. This results in 1000 catalogs each containing on average $\sim\!1.1\times10^5$ individual events with total source-frame masses ranging from $\sim\!7\,{\rm M}_\odot$ to $\sim\!600\,{\rm M}_\odot$.
The sky localization, inclination, polarization angle, time and phase of coalescence, $\phi_{12}$, and $\phi_{\rm JL}$ are sampled from uniform, non-informative distributions. We then compute the observability of the sources with the future detector networks described in Sec.~\ref{subsec:detectors}. We also carry out a separate analysis for the maximum-likelihood sample.

\begin{figure}[tb]
    \centering
    \includegraphics[width=\linewidth]{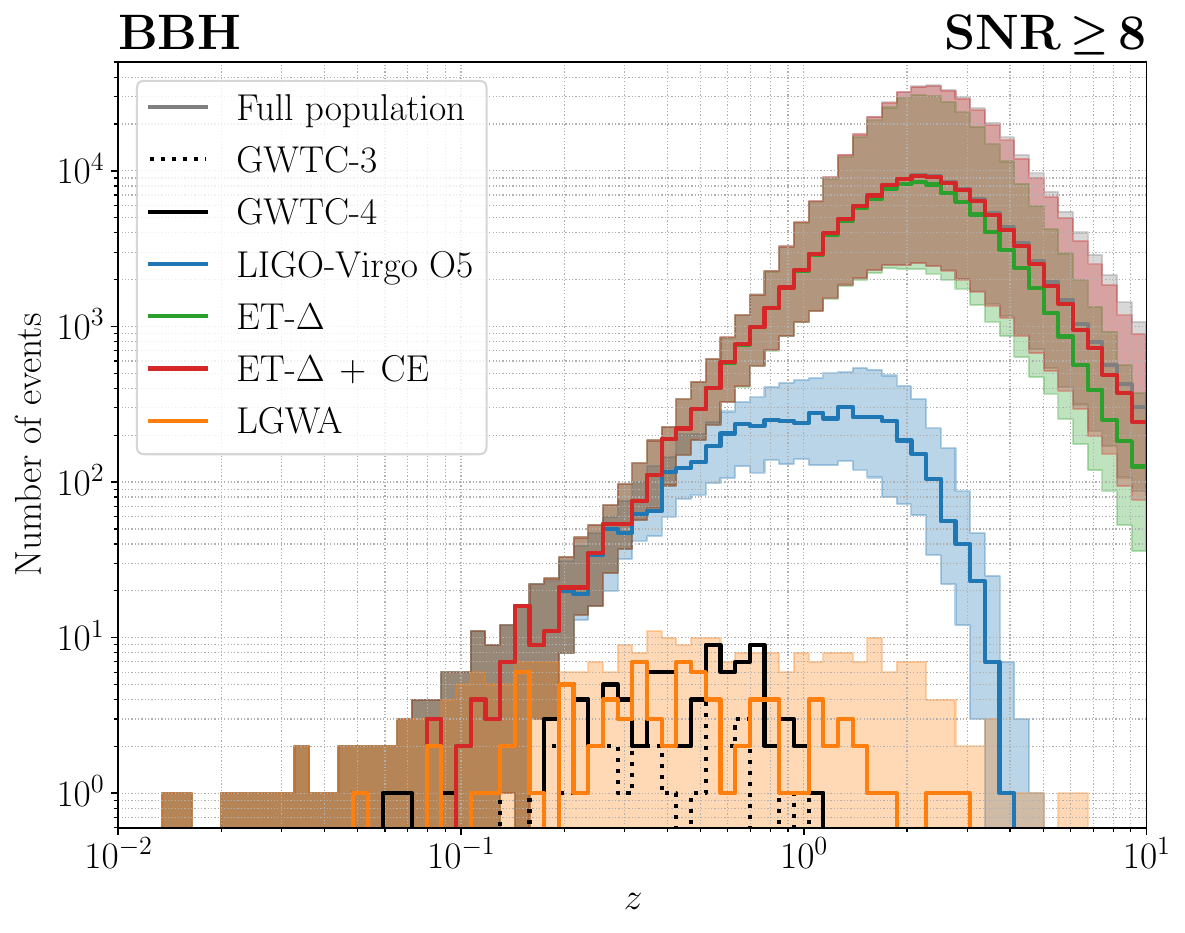}
    \caption{Histogram of the redshift distribution of the simulated catalog for events detected with ${\rm SNR}\geq8$ using the different detectors and networks considered in this work. The solid lines report the population corresponding to the maximum likelihood estimated in Ref.~\cite{LIGOScientific:2025pvj} for the reference population model, while the shaded bands enclose the maximum and minimum value across 1000 simulated catalogs with varying hyperparameters. For reference, we further report the distribution of maximum-likelihood redshifts for the sources in GWTC-3~\cite{KAGRA:2021vkt} and GWTC-4~\cite{LIGOScientific:2025slb}.}
    \label{fig:zhist_detections}
\end{figure}

In Fig.~\ref{fig:cumulative_SNR_gwtcpop} we show the resulting cumulative SNR distribution of the sources, while in Fig.~\ref{fig:zhist_detections} we report the redshift distribution of the sources detected with ${\rm SNR}\geq 8$.
We find that, even operating alone, LGWA will be able to observe $87\substack{+44\\-46}$ sources with ${\rm SNR}\geq 8$ out to $z\sim2.8$, and $405\substack{+131\\-152}$ with ${\rm SNR}\geq 5$ out to $z\sim4.9$. We remark again that this is just the number of sources accessible to LGWA which will merge in the ground-based detectors band. A population of more massive IMBH binaries observable only with lower frequency detectors would have different detection rates, but given the large astrophysical uncertainties on IMBHs, we prefer to focus on the mass range already constrained by LVK observations. 

\begin{figure*}[htb]
    \begin{tabular}{cc}
        \includegraphics[width=0.49\linewidth]{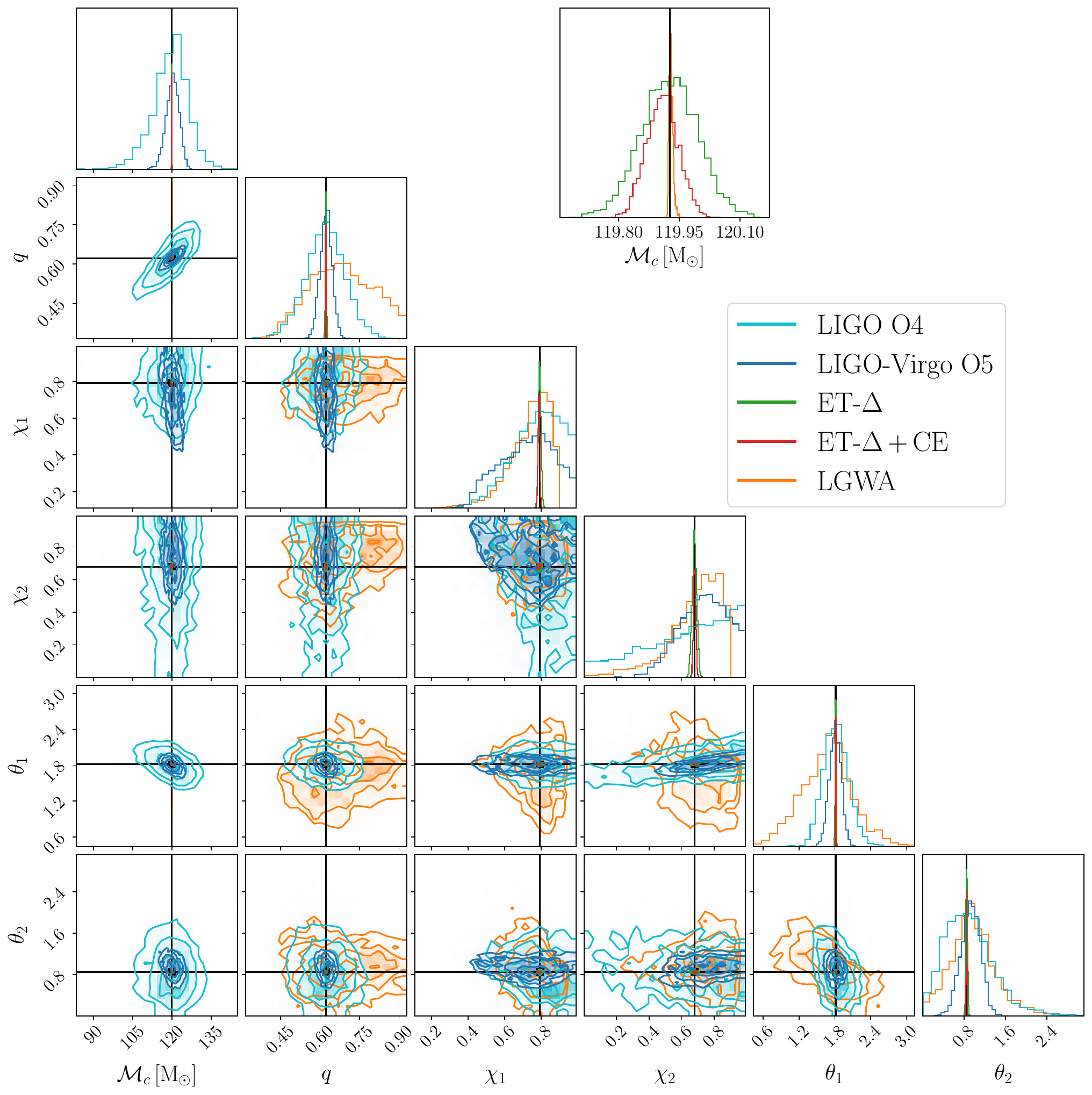} & \includegraphics[width=0.49\linewidth]{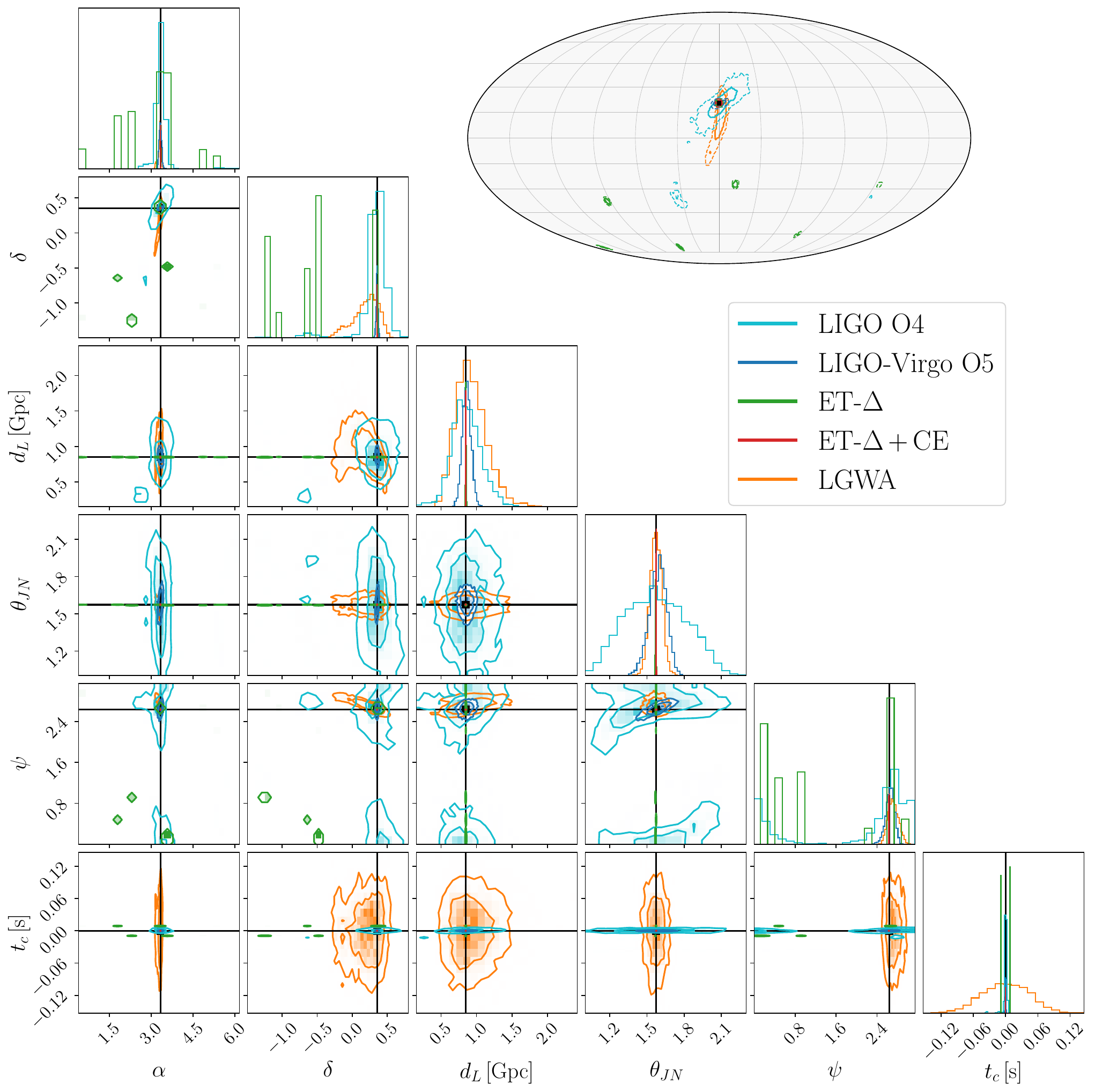}\\
    \end{tabular}
    \caption{Results for our injection study of a GW231123-like event at the various networks considered in this work. We also report the posterior released by LVK for the \textsc{IMRPhenomXPHM}, the same one used in our analysis. In the right panel we show the results for some intrinsic parameters: chirp mass, mass ratio $q=m_2/m_1$, and spin magnitudes and tilts of the two objects. The inset zooms in the marginal ${\cal M}_c$ distribution. In the right panel we show the results for some extrinsic parameters: sky position, luminosity distance, inclination, polarization angle and time of coalescence (the latter centered on the injection value). The inset zooms in on the sky position distribution, with a right ascension reference centered on the injection value. 
    }
    \label{fig:corners_231123}
\end{figure*}

For ground-based detectors, we find that a LIGO-Virgo network at design sensitivity would be able to detect $4481\substack{+2838\\-1984}$ binaries per year with ${\rm SNR}\geq 8$, among which $83\substack{+27\\-42}$ ($382\substack{+112\\-138}$) would be detectable by LGWA with ${\rm SNR}\geq 8$ (${\rm SNR}\geq 5$), and $1534\substack{+718\\-521}$ would have a joint SNR above 12. This is an improvement with respect to the $1393\substack{+610\\-456}$ detections with SNR higher than 12 in the LIGO-Virgo network alone. 

Moving to 3G detectors, ET alone will be able to deliver an impressive $7.9\substack{+16\\-5.2} \times 10^4$ detections with ${\rm SNR}\geq 12$. For a network including CE this number would raise to $1.1\substack{+2.4\\-0.7} \times 10^5$. In both cases, 3G detectors would observe all of the events that are detectable by LGWA. Crucially, the BBH signals observable in the deci-Hz band merge in the ground-based detector band after short time delays, at most $\sim\!1$\,yr across all simulated binaries. This is very different from milli-Hz multi-band observations: LISA binaries sometimes reach the Hz band after $\gtrsim{\cal O}(10\,{\rm yr})$, consequently lowering the joint detection rates (see e.g.~\cite{Gerosa:2019dbe}). This can be also appreciated from Fig.~\ref{fig:psds_signals}, in which each of the reported signals is truncated 10\,yr before merger.

{%
\renewcommand{\arraystretch}{1.2}
\setlength{\tabcolsep}{14pt}
\begin{table}[b]
    \centering
    \begin{tabular}{lcc}
        \toprule\midrule
        \multicolumn{3}{c}{\textbf{Massive BBHs}} \\ 
        \midrule\midrule
        \multirow{2}{*}{Detector} & Number of & Average number\\
        & detections & of cycles \\
        \midrule
        LVK O5 & $24\substack{+90\\-23}$ & 3 \\
        ET-$\Delta$ & $87\substack{+363\\-80}$ & 41 \\
        ET-$\Delta\,+\,$CE & $87\substack{+364\\-80}$ & 41 \\
        LGWA & $9\substack{+50\\-9}$ & $ 1.5 \times 10^5$\\
        \midrule
        Total events & $87\substack{+365\\-80}$ \\
        \midrule\bottomrule
    \end{tabular}
    \caption{Number of detections with ${\rm SNR}\geq8$ for binaries with primary source-frame mass $m_{1,\,s}\geq100\,{\rm M}_\odot$, and for the different detectors considered in this work. In the third column we list the average number of cycles spent by these binaries in the sensitive band of each instrument. The last row reports the total number of events with $m_{1,\,s}\geq100\,{\rm M}_\odot$.}
    \label{tab:massive_bbh}
\end{table}
}%

Let us focus on the high-mass range of the simulated catalogs, that most benefits from a low-frequency detector. In Table~\ref{tab:massive_bbh} we report the total number of BBHs with a primary source-frame mass $m_{1,\, s}\geq100\,{\rm M}_\odot$, and the number of such high-mass binaries detectable by the various detector networks. 
The number of simulated sources and the number of detections with the various networks both fluctuate significantly with the catalog realization, ranging from a few to a few hundreds, because of the assumed power-law mass model. In general, we find that LGWA alone can detect a few to a few tens of events with ${\rm SNR}\geq 8$, while the LIGO-Virgo O5 run can detect about twice as many. Conversely, 3G detectors will be able to observe almost all binaries in this mass range. It is interesting to compare the number of cycles spent by these binaries in the detector band before merger. On average, these massive BBH systems spend only a few inspiral cycles in 2G detectors. The number of cycles grows by about an order of magnitude in 3G detectors (which have better low-frequency sensitivity). Quite remarkably, the number of cycles spent by these binaries in the sensitivity band of a deci-Hz observatory can be as high as ${\cal O}(10^5)$. This has interesting consequences in the context of PE, as we shall see below.

\subsection{Parameter estimation for GW231123}\label{subsec:pe_231123}

We now quantify the PE capabilities of LGWA by focusing on a specific example: a massive, GW231123-like event. We perform a zero-noise injection with the parameters listed in Table~\ref{tab:231123_params_and_priors} in each detector network (see Sec.~\ref{subsec:data} for details about the priors and sampler). For ground-based detectors, we find the injection to have an optimal matched-filter SNR of 75.8 in the LIGO-Virgo O5 network, of 581.4 in ET-$\Delta$, and of 1249.8 in ET-$\Delta$+CE. The SNR in LGWA with our choice of position and orientation is more modest (11.9), while the SNR for the actual event detected by LIGO in O4 is 22.6.

The results of our injection study are reported in Fig.~\ref{fig:corners_231123}. In the left panel we focus on the binary's intrinsic parameters, while in the right panel we consider the extrinsic parameters. These results allow us to draw some interesting conclusions.

Despite the modest SNR, the LGWA achieves impressive accuracy in the estimation of the detector-frame chirp mass, doing better than the ET-$\Delta$+CE network (see the top-right inset in the left panel). This is because the system spends an extremely large number of cycles in band, $\sim\!10^5$, and the relative uncertainty on ${\cal M}_c$ scales as the inverse of the number of cycles~\cite{Maggiore:2007ulw}. The relative standard deviation of the samples $\sigma_{{\cal M}_c}/{\cal M}_c\sim\!3.4\times10^{-5}$ is indeed of the same order of magnitude. The difference can be partly understood from the fact that the SNR in each cycle is not the same, and some cycles are at very low SNR, resulting in a broader posterior.

The binary has an LGWA SNR of 11.9 (much smaller than its LIGO O4 SNR of 22.6), but the spin magnitudes are estimated with comparable accuracy. In fact, the spin magnitude accuracy in LGWA is even comparable to the LIGO-Virgo O5 network (with an SNR of 75.8). The best accuracy on the spin magnitudes is achieved by 3G detector networks, where the SNR is in the hundreds. Concerning the precessing spin parameters, the accuracy achievable in the reconstruction of the spin tilts with LGWA is comparable to that attainable with current instruments, while $\phi_{\rm JL}$ and $\phi_{12}$ remain unconstrained. This can again be traced to the low SNR of the source in LGWA, with most of the cycles being too faint to observe the imprint of spin precession.

As for the extrinsic parameters, the LGWA as a single instrument has slightly better angular localization than the two LIGO detectors in the ongoing O4 run. Besides, $\alpha$ and $\delta$ exhibit a different correlation (see the top-right inset in the right panel). This accuracy is again due to the length of the signal in band, which allows the detector to ``self-triangulate'' with good accuracy.
As discussed e.g. in Refs.~\cite{Baibhav:2020tma,Singh:2020wsy,Santoliquido:2025lot}, for short signals the angular resolution of a triangular detector such as ET shows multimodalities, despite the high SNR. For such systems, the LGWA sky position estimate allows us to select the correct mode in the sky. A full multiband analysis including phase coherence between the different detectors would further improve sky localization~\cite{Wu:2025zhc}.
Note that the sky localization is very accurate despite the relatively poor LGWA estimate of the time of coalescence, $t_{\rm GPS}$, compared to ground-based instruments. The time of coalescence is hard to estimate because LGWA operates at low frequency and it does not directly observe the merger, but this inaccuracy is partially compensated by the long observing time in the deci-Hz band.

The posterior for the inclination angle $\theta_{\rm JN}$ in the two LIGO O4 detectors is multimodal. This is because the two detectors are aligned and ``see'' the same mixture of the two polarizations in the signal. The multimodality is indeed broken once we add Virgo to the network, or if we consider 3G interferometers. Despite operating in isolation, LGWA is still capable of recovering a unimodal posterior for $\theta_{\rm JN}$. This is again because the signal spends a long time in band, which allows the detector to move during the observation and to disentangle the contribution of the $+$ and $\times$ polarizations.

Concerning ground-based detectors, it is clear that 3G instruments yield exquisite PE accuracy for all source parameters. In particular, a single ET detector does not do much worse than the ET+CE network in the reconstruction of the intrinsic parameters, despite the lower SNR.

\section{Conclusions}

Massive BBHs are arguably among the most interesting astrophysical systems observable though GWs. These binaries merge at frequencies $\lesssim{\cal O}(100\,{\rm Hz})$, and as such they are observable with both present and future ground-based interferometers, but much longer observation times of these signals are possible at frequencies below-Hz. For this reason, in this paper we investigate the prospects of observing massive BBHs with the proposed LGWA deci-Hz detector. 

We estimate that $\sim\!30\%$ of the events in the public LVK catalog would be observable by LGWA with ${\rm SNR}\geq 8$, and $\sim\!57\%$ could be found with an LGWA ${\rm SNR}\geq 5$ in archival data. By simulating different realizations of the LVK population, we estimate that the LGWA alone could detect between few tens (and up to more than a hundred) of the binaries merging in the band accessible to ground-based detectors. Remarkably,  {\em a few hundreds} of these events can be used for multiband observations out to high redshift, $z\gtrsim4$. 
This highlights the high complementarity between LGWA and present or future ground-based detectors.

Deci-Hz detectors can observe a very large number of cycles, and they are especially useful to estimate the parameters of high-mass BBH mergers. Our injection study of a GW231123-like event shows that they can estimate the chirp mass even better than ET and CE (where the SNR is two orders of magnitude larger than the LGWA SNR). Moreover, we have shown that the long signal duration implies that even a single lunar instrument can ``self-triangulate'' to obtain accurate sky localization and disentangle the signal polarizations, leading to tight inclination posteriors. These features are complementary to ground-based observations. A coherent analysis will further improve constraints on masses, spins, sky position, and timing beyond what detectors in either band can achieve alone.

Our forecasts rely on current population models (given astrophysical uncertainties on IMBHs), and they do not yet include realistic duty cycles and data gaps (see e.g.~\cite{Burke:2025bun} for a discussion in the context of LISA), calibration, or waveform systematics. It will be crucial to addressing all of these issues, to develop a fully coherent multiband pipeline. It is also important to better understand the lunar environment. Even with these limitations, the science case is clear: opening the deci-Hz band will transform multiband GW astronomy, improving our reconstruction of the BBH mass spectrum and enabling more precise inference for massive BBHs. %

\begin{acknowledgments}
We thank Michele Mancarella, Konstantinos Kritos, Luca Reali, Jay Wadekar,  and Akshita Mittal for interesting discussions and comments. 
F.I. and E.B. are supported by NSF Grants No.~AST-2307146, No.~PHY-2513337, No.~PHY-090003, and No.~PHY-20043, by NASA Grant No.~21-ATP21-0010, by John Templeton Foundation Grant No.~62840, by the Simons Foundation [MPS-SIP-00001698, E.B.], by the Simons Foundation International [SFI-MPS-BH-00012593-02], and by Italian Ministry of Foreign Affairs and International Cooperation Grant No.~PGR01167.
This work was carried out at the Advanced Research Computing at Hopkins (ARCH) core facility (\url{https://www.arch.jhu.edu/}), which is supported by the NSF Grant No. OAC-1920103. 
The work of F.I. is supported by a Miller Postdoctoral Fellowship. 
We acknowledge financial support from the Italian Space Agency (ASI) under Grant No. 2025-29-HH.0.

\textit{Data Availability Statement:} The generated data, posterior samples, and analysis scripts used in this work are publicly available on \textsc{Zenodo}~\raisebox{-1pt}{\href{https://doi.org/10.5281/zenodo.19716427}{\includegraphics[width=9pt]{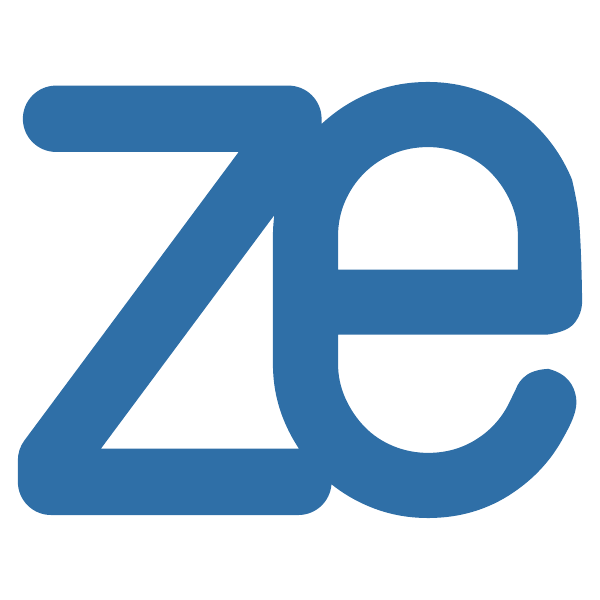}}}~\cite{OurpaperDataPub}.

\end{acknowledgments}

\bibliography{lgwa231123}

\end{document}